\documentclass[letterpaper]{article} 
\usepackage{aaai23}  
\usepackage{times}  
\usepackage{helvet}  
\usepackage{courier}  
\usepackage[hyphens]{url}  
\usepackage{graphicx} 
\usepackage{natbib}  
\usepackage{caption} 
\usepackage{algorithm}
\usepackage{algorithmic}

\usepackage{multirow}
\usepackage{amsmath}
\usepackage{amssymb}

\usepackage{newfloat}
\usepackage{listings}
\DeclareCaptionStyle{ruled}{labelfont=normalfont,labelsep=colon,strut=off} 
\floatstyle{ruled}
\newfloat{listing}{tb}{lst}{}
\floatname{listing}{Listing}
\title{Multi-Modality Deep Network for Extreme Learned Image Compression}

\author{
		Xuhao Jiang\textsuperscript{\rm 1}, Weimin Tan\textsuperscript{\rm 1}, Tian Tan\textsuperscript{\rm 1}, Bo Yan\textsuperscript{\rm 1}\thanks{Corresponding author: Bo Yan. This work is supported by NSFC (Grant No.: U2001209, 61902076) and Natural Science Foundation of Shanghai (21ZR1406600).}, Liquan Shen\textsuperscript{\rm 2}	
}
\affiliations{
	\textsuperscript{\rm 1}School of Computer Science, Shanghai Key Laboratory of Intelligent Information Processing, Shanghai Collaborative Innovation Center of Intelligent Visual Computing, Fudan University, Shanghai, China\\
	\textsuperscript{\rm 2}School of Communication, Shanghai University, Shanghai, China\\
	20110240011@fudan.edu.cn, wmtan@fudan.edu.cn, tant21@m.fudan.edu.cn, byan@fudan.edu.cn, jsslq@163.com
}

\usepackage{bibentry}

\begin{document}

\maketitle

\begin{abstract}
Image-based single-modality compression learning approaches have demonstrated exceptionally powerful encoding and decoding capabilities in the past few years , but suffer from blur and severe semantics loss at extremely low bitrates. To address this issue, we propose a multimodal machine learning method for text-guided image compression, in which the semantic information of text is used as prior information to guide image compression for better compression performance. We fully study the role of text description in different components of the codec, and demonstrate its effectiveness. In addition, we adopt the image-text attention module and image-request complement module to better fuse image and text features, and propose an improved multimodal semantic-consistent loss to produce semantically complete reconstructions. Extensive experiments, including a user study, prove that our method can obtain visually pleasing results at extremely low bitrates, and achieves a comparable or even better performance than state-of-the-art methods, even though these methods are at 2$\times$ to 4$\times$ bitrates of ours.

\end{abstract}

\section{Introduction}
\label{sec:intro}

During the past decades, image data on the Internet shows an explosive growth, bringing huge challenges for data storage and transmission. To meet this ever-increasing requirements, low-bitrate lossy image compression is a promising way to save storage and transmission bandwidth. Traditional image compression algorithms, e.g., Better Portable Graphics (BPG)~\cite{BPG} and Versatile Video Coding (VVC)~\cite{VVC}, are widely used in practice. However, they will cause serious blocking artifacts due to block-based processing at low bitrates. Therefore, exploring better methods for extreme image compression is urgently needed.

\begin{figure}[t]
	\begin{center}
		\includegraphics[width=1\linewidth]{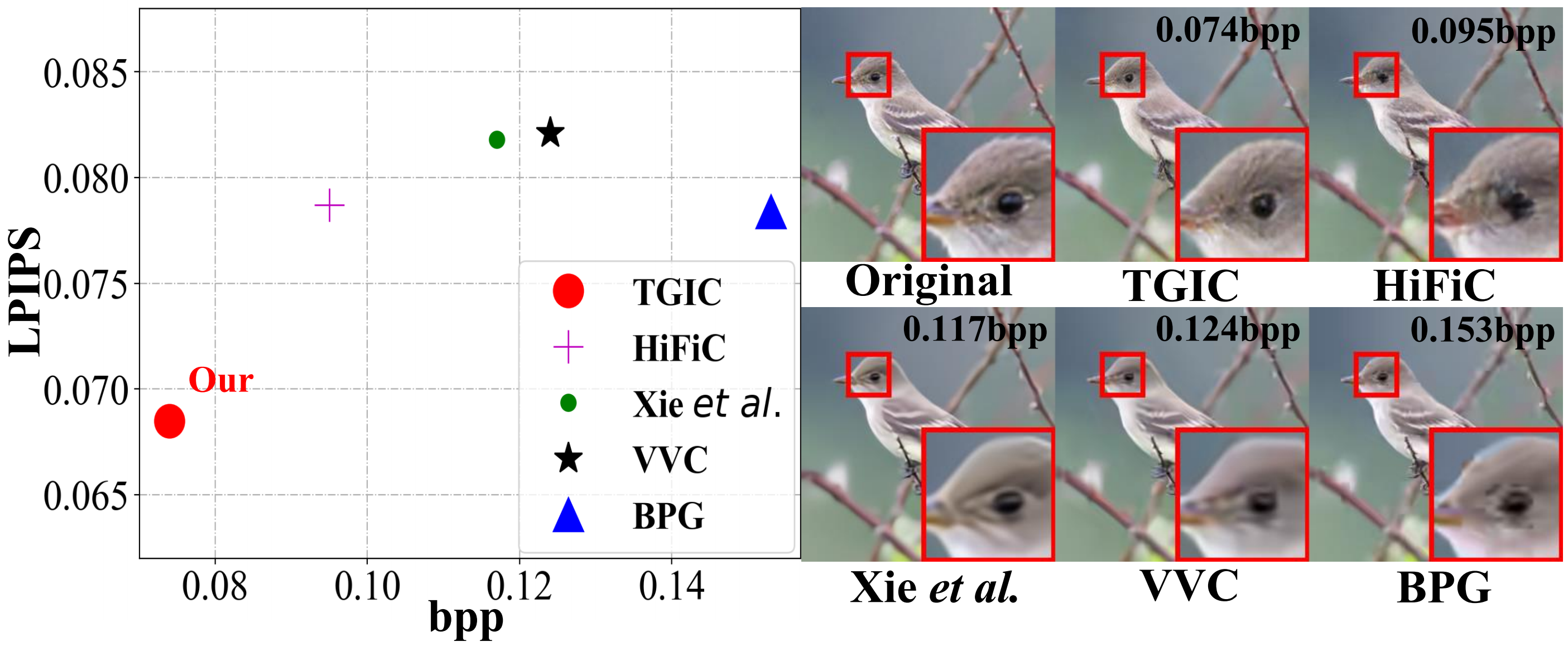}
	\end{center}
	\caption{Visual comparisons of the proposed TGIC and some state-of-the-art methods (HiFiC~\cite{mentzer2020high}, VVC~\cite{VVC}, BPG~\cite{BPG} and the work of Xie \emph{et al.}~\cite{xie2021enhanced}).}
	\label{fig:shouye}
\end{figure}

Recently, many single-modality learned methods~\cite{xie2021enhanced,mentzer2020high} have been proposed. However, they also fail to reconstruct satisfactory results at extremely low bitrates. Specifically, they may generate blurry results due to limited bits, or utilize Generative Adversarial Networks (GAN) to produce sharp results whose textures may not be semantically consistent with the original image, as shown in Fig.~\ref{fig:shouye}. The text-to-image synthesis task is currently receiving a lot of attention, which generates semantically consistent images from text descriptions. Inspired by this task, multi-modality image compression may have great advantages at low bitrates. The corresponding text provides the high-level image semantic information, which can be used as prior information to assist image compression. Specifically, the text describes a rough content of the image and its local features, such as, color, location, shape, etc. This semantic information can be used to assist in reconstructing images, which can help save image bits. In addition, under the guidance of text semantic information, the multi-modality model can generate visually pleasing results that are more semantically consistent with the original image, as shown in Fig.~\ref{fig:shouye}. Note that the text description occupies very few bits and can be transmitted to the decoder side at marginal bandwidth cost. Even for low bitrates (0.2 bpp), text uses less than one-twentieth of the bits used by images on datasets~\cite{wah2011caltech,nilsback2008automated}.

In this paper, a text-guided image compression (TGIC) generative adversarial network is proposed, in which the text description is utilized as prior information to assist in image compression. TGIC can produce better results compared with other methods, even if we use a much lower bitrate. For the image encoding, based on the image-text attention (ITA) module, text information is introduced into the codec to guide the generation of compact feature representations. In the image decoding stage, we design an image-request complement (IRC) module to adaptive fuse the text and image information for better reconstructions. Besides, an improved multimodal semantic-consistent loss is designed to further improve the perceptual quality of reconstructions. The main contributions are as follows:
\begin{itemize}
	\item We propose a novel codec framework for image compression, which utilizes the semantic information of the text description to improve coding performance. To the best of our knowledge, this is the first attempt that uses text semantic information as prior information to guide image compression.

	\item  We fully study the role of text description in different components of the codec, and demonstrate its effectiveness for image compression. In particular, we adopt ITA to fuse image and text features, and propose IRC that allows the network to adaptively learn the much-needed guidance knowledge from text.

	\item The experiments (including a user study) show the outstanding perceptual performance of our TGIC in comparison with the existing learned image compression methods and traditional compression codecs.

\end{itemize}


\section{Related Work}
\subsection{Lossy Image Compression}
Lossy image compression has received significant attention from both academia and industry due to its huge practical value. Traditional compression standards, such as JPEG~\cite{wallace1992jpeg}, JPEG2000~\cite{rabbani2002jpeg2000}, BPG (HEVC-Intra)~\cite{BPG} and VVC~\cite{VVC} (the latest traditional codec), reply on hand-crafted module design. However, they ignore spatial correlations between image blocks, which results in image discontinuities at block boundaries.

Recently, many learned methods have been proposed to tackle the problem of image compression, and achieve promising results. Some early methods~\cite{toderici2015variable,toderici2017full} utilize the recurrent neural network to recursively compress the residual information, but they cannot directly optimize the rate in the training phase. The subsequent works are mainly based on variational autoencoder, and significant advances have been made progressively~\cite{rippel2017real,zhang2019residual,chen2021end,balle2017end,agustsson2017soft,theis2017lossy,balle2018variational,mentzer2018conditional,lee2018context,hu2021learning,cheng2020learned}. Recognizing the huge potential of the hyperprior model~\cite{balle2018variational}, many follow-up methods improve the entropy estimation techniques based on hyperprior design, such as coarse-to-fine model~\cite{hu2020coarse}, joint model~\cite{minnen2018joint} and 3D context entropy model~\cite{guo20203}. Besides, some methods~\cite{agustsson2019generative,mentzer2020high,tschannen2019deep} based on GAN~\cite{goodfellow2014generative} have been proposed for image compression at low bitrates. Mentzer \emph{et al.}~\cite{mentzer2020high} investigate normalization layers, generator and discriminator architectures, training strategies, as well as perceptual losses, and propose HiFiC which shows impressive performance.

However, the above-mentioned algorithms all show poor performance at extremely low bitrates, and even the excellent GAN-based method HiFiC shows general performance in this situation. The main reason is that it is impossible to faithfully reconstruct the entire content of the uncompressed image with extreme limited bits, and the GAN-based methods cannot generate images with realistic textures due to the lack of guidance from additional prior information. Therefore, benefiting from the semantic information provided by the text description, image compression based on multimodal machine learning may have a greater possibility to obtain better compression performance.

\subsection{Multimodal Machine Learning}

The multimodal machine learning has recently become a very hot topic due to its powerful advantages in the field of computer vision, such as text-to-image synthesis and image captioning. The text-to-image synthesis task aims to produce a high-quality image from a described text, such as~\cite{zhang2017stackgan,xu2018attngan,reed2016generative}. For example, given the text description, AttnGAN~\cite{xu2018attngan} employs attention mechanism to produce images with photorealistic details. Contrary to the text-to-image synthesis task, the image captioning task~\cite{shi2020improving,anderson2018bottom,xu2015show} is to generate a corresponding text description for a given image. Inspired by these works, we propose text-guided image compression, which uses the semantic information of text description as prior information to improve coding performance.

\section{Method}
\begin{figure}[t]
	\begin{center}
		\includegraphics[width=1\linewidth]{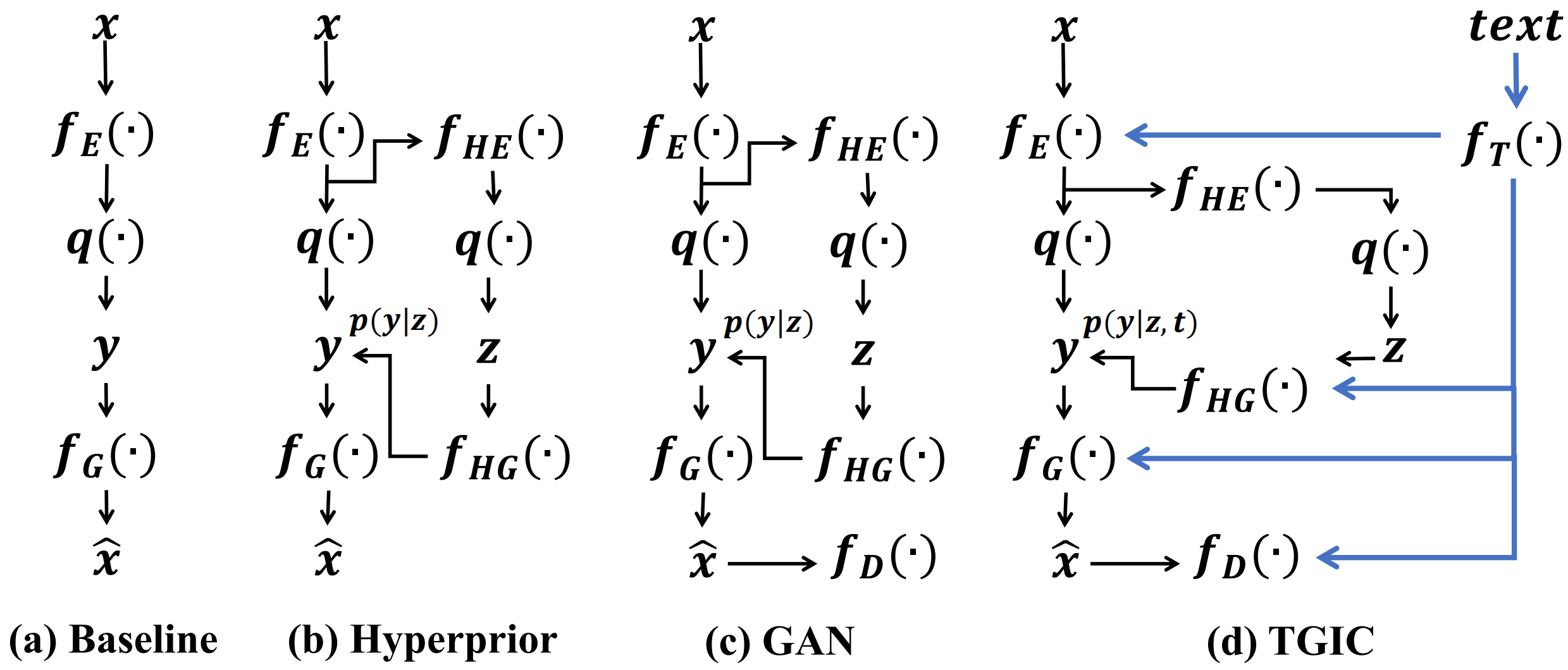}
	\end{center}
	\caption{Overview of existing learned compression methods and the proposed TGIC. $f_{E}(\cdot)$ and $f_{G}(\cdot)$ represent the encoder and decoder of codec, respectively, and $q(\cdot)$ represents the quantization function. $f_{HE}(\cdot)$ and $f_{HG}(\cdot)$ denote the encoder and decoder of the auxiliary autoencoder (entropy model), respectively. $f_{D}(\cdot)$ and $f_{T}(\cdot)$ denote the discriminator and the text encoder, respectively.}
	\label{fig:methods-class}
\end{figure}

\begin{figure*}[h]
	\begin{center}
		\includegraphics[width=1\linewidth]{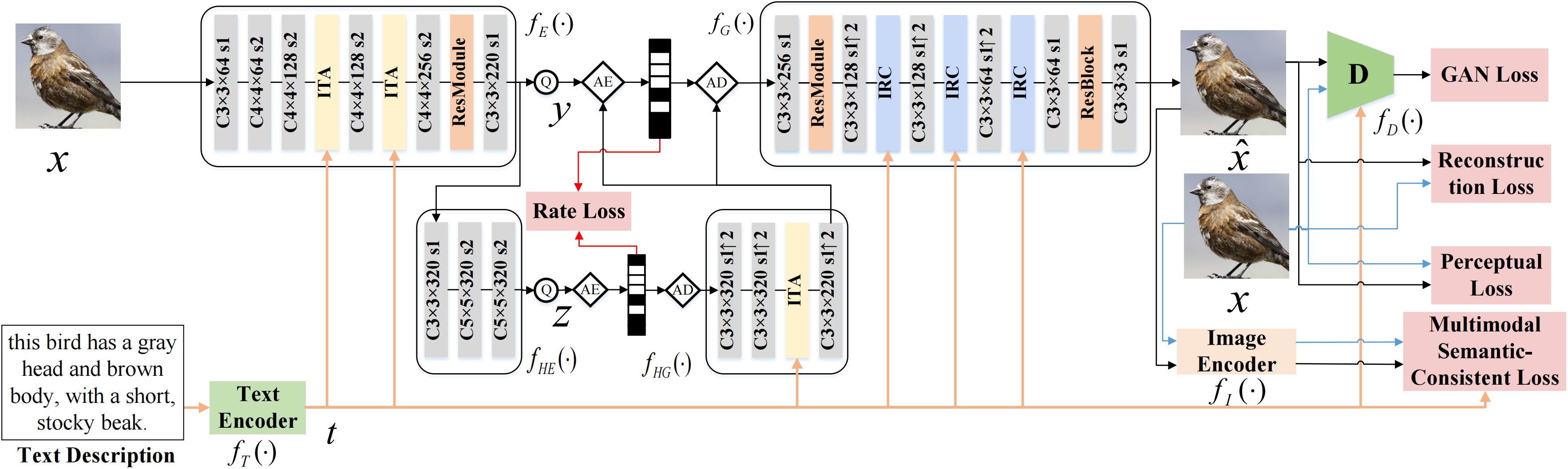}
	\end{center}
	\caption{Architecture of our text-guided image compression (TGIC) model. $C3\times3\times64~s2$ is a convolution with 64 channels, with $3\times3$ filters and stride 2. $\uparrow2$ indicates the nearest neighbor upsampling. In addition, AE and AD are arithmetic encoding and decoding, and Q is for quantization. ITA is introduced to fuse image features and text features based on attention mechanism, and IRC is designed to adaptively use the text features for the image semantic complement. Resblock and ResModule are based on ~\cite{he2016deep}.}
	\label{framework}
\end{figure*}

\subsection{Preliminaries}

Figure. \ref{fig:methods-class} provides an overview of learned image compression in the transform coding approach~\cite{goyal2001theoretical}. As shown in Fig.~\ref{fig:methods-class} (a), the baseline model can be expressed as
\begin{equation}
	y=q(f_{E}(x)),~then~\hat{x}=f_{G}(y),
\end{equation}
where $x$, $\hat{x}$ and $y$ are raw images, reconstructed images and compressed codes, respectively. Specifically, $x$ is encoded by an encoder $f_{E}(\cdot)$, and then quantized by $q(\cdot)$ to obtain $y$. $y$ is then losslessly compressed into a bitstream by using entropy coding like arithmetic coding~\cite{rissanen1981universal}. For decoding, the decoder $f_{G}(\cdot)$ transforms $y$ to obtain the $\hat{x}$. Here, the widely used quantization function proposed by \cite{balle2017end} is employed in our method. The corresponding rate $R$ of $y$ is estimated by a fully factorized density model $p_{y|\theta}$ during training, which is formulated by
\begin{equation}
	R= \mathbb{E}[-log_{2}p_{y|\theta}(y|\theta)].
\end{equation}

Balle \emph{et al.} first proposes a hyperprior model~\cite{balle2018variational}, which utilizes the side information $z$ to capture the spatial dependence of $y$, as shown in Fig.~\ref{fig:methods-class} (b). This idea is realized by introducing an additional entropy model. The side information $z$ can be calculated by $z=q(f_{HE}(y))$, where $f_{HE}(\cdot)$ denotes the encoder of the entropy model. Then $z$ is transformed by the decoder of the auxiliary autoencoder $f_{HD}(\cdot)$, which is used to estimate the distribution of $y$. Then the rate $R$ is formulated by
\begin{equation}
	R= \mathbb{E}[-log_{2}p_{y|z}(y|z)]+\mathbb{E}[-log_{2}p_{z|\theta}(z|\theta)].
\end{equation}
As shown in Fig.~\ref{fig:methods-class} (c), compared with the hyperprior model, the GAN-based model has one more discriminator, which generates the reconstructed images with fine-grained details through adversarial training.

Based on the previous works, we propose a new codec framework (as shown in Fig.~\ref{fig:methods-class} (d)), which utilizes the text information to assist in image compression. We will describe the proposed framework in detail in the following section.

\begin{figure}[t]
	\begin{center}
		\includegraphics[width=0.9\linewidth]{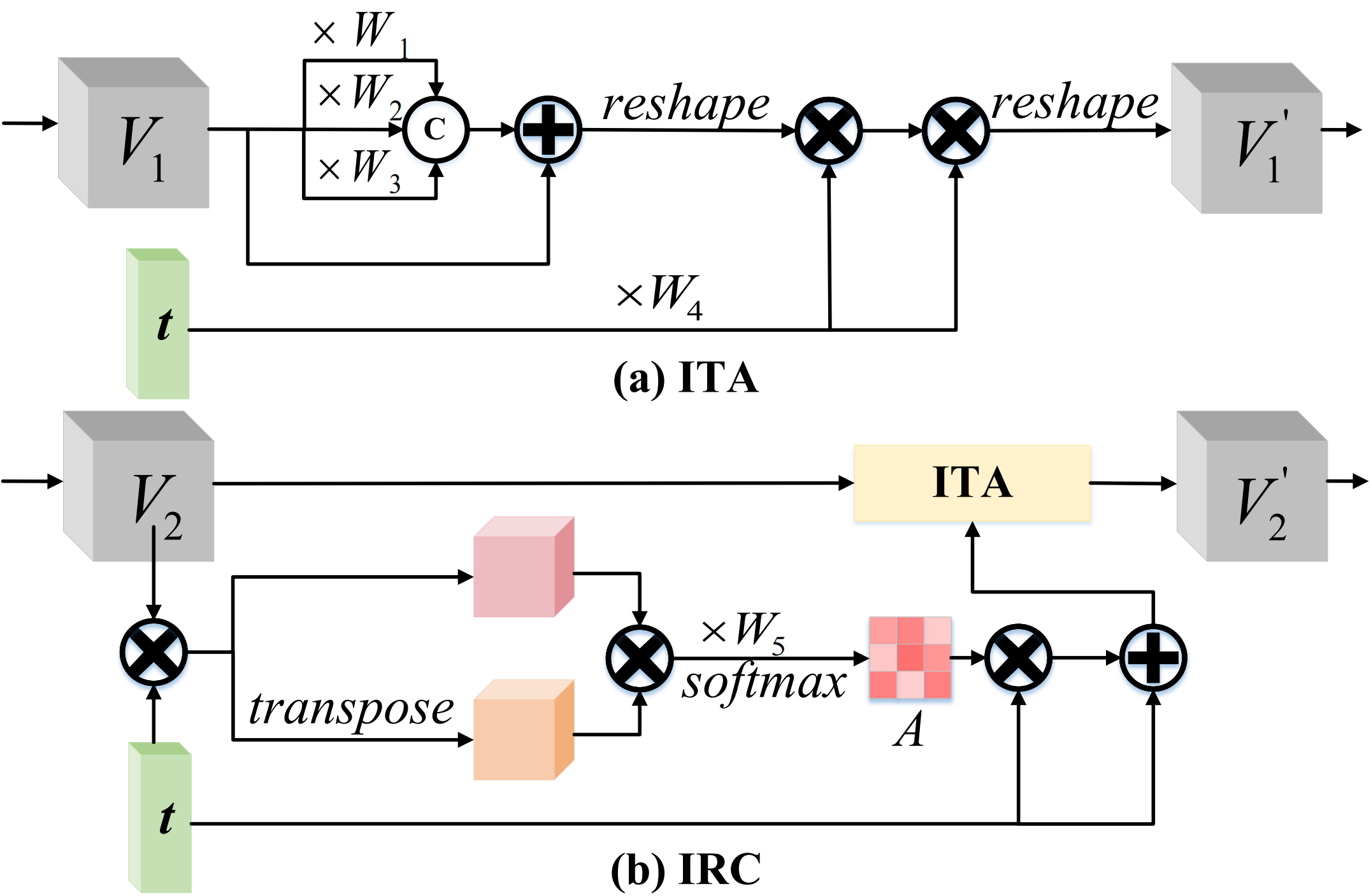}
	\end{center}
	\caption{Architectures of ITA and IRC. C means concatenation. $V_1$, $V_2$, $V_1'$ and $V_2'$ are the image features.}
	\label{IRC}
\end{figure}

\subsection{Proposed Method}

The architecture design of TGIC is shown in Fig. \ref{framework}, which describes the details of the Fig. \ref{fig:methods-class} (d). Specifically, the main body of TGIC is composed of four components: encoder,  decoder, entropy model and discriminator. The image encoder and text encoder are employed to map the image and the text into a common semantic feature space. Firstly, the text features are extracted from the text description by using the text encoder. Then these text features are introduced into TGIC to assist image compression for better image reconstruction based on the image-text attention (ITA) module. In particular, we design an image-request complement module (IRC) in the decoder, so as to realize adaptive selection of text semantic information for the image semantic feature enhancement. Besides, we also design an improved multimodal semantic-consistency loss, which considers the semantic consistency between the reconstructions and the texts, as well as the uncompressed images.

\subsection{Text-Guided Feature Representation }

In our TGIC, the text encoder is a bi-direction Long Short-Term Memory (LSTM)~\cite{schuster1997bidirectional}, which extracts the semantic features of texts. The corresponding calculation can be defined as
\begin{equation}
	t=f_T(text),
\end{equation}
where $t$ denotes the text semantic features and $f_T(\cdot)$ represents the text encoder. Then $t$ is input to the encoder and the entropy model respectively to achieve a compact feature representation.

\textbf{Text guidance in the encoder.}
ITA is adopted to calculate the correlation between the text and image features, and then fuse these two features, as shown in Fig.~\ref{IRC} (a). Inspired by \cite{xu2018attngan,li2021perceptual}, ITA uses multi-scale residual structure to further extract image features, and use matrix multiplication to calculate correlation, where $W_1$, $W_2$ and $W_3$ represent the convolutional operations with different filter size, and $W_4$ is used to adjust the text feature dimension. The previous works~\cite{agustsson2019generative,duan2020jpad} have verified that the semantic segmentation maps can improve coding performance. Considering that text can provide high-level semantic information of images, we utilize the high correlation between image and text features to achieve more compact image features. Therefore, $y$ can be defined as $y=q(f_E(x,t))$, where $f_E(\cdot)$ represents the encoder of TGIC.

\textbf{Text guidance in the entropy model.} The hyperprior model~\cite{balle2018variational} improves the coding performance by introducing the side information, whose essence lies in predicting the distribution of latent features. The distribution of latent features is closely related to the content of the image. Considering that the text description can offer the image semantic information, the text information may help to predict the distribution of latent features. Inspired by \cite{li2021deep}, the text features are introduced to $f_{HG}(\cdot)$ of the entropy model. Similarly, this operation is also based on ITA. Then the entropy model can use the semantic relevance of text and image to predict the distribution of latent features more accurately. Under the guidance of the text features, the estimated distribution is converted from $p_{y|z}$ to $p_{y|z,t}$, which better parameterizes the distributions of latent codes and improves the entropy model performance.

\subsection{Text-Guided Image Reconstruction}
Due to the quantization operation, the image features in the decoder  inevitably lose some information. Considering that the text description contains some semantic information of the image, the semantic features of the text is considered as prior information to enhance the image features. Then, the reconstructed image can be defined as $\hat{x}=f_G(y, t)$, where $f_G(\cdot)$ represents the decoder of TGIC. Aiming to make better use of text features for image feature enhancement, we propose the IRC to adaptively fuse the text information and image information. The architecture of IRC is shown in Fig. \ref{IRC} (b). Firstly, IRC predict an attention map $A$ based on the correlation between the input image features $V_2$ and text features, which is realized by the matrix multiplication of feature maps and is calculated as
\begin{equation}
	A=softmax(W_5((V_2t)(V_2t)^T))
\end{equation}
where $W_5$ represents the convolutional operation. Then we use the obtained $A$ to weight and add the text features. Finally, the adaptive selected text features and the input image features are fused to obtain the enhanced features $V_2'$ by using ITA. This process can be expressed as $V_2'=ITA(V_2,At+t)$.

\subsection{Text-Guided Adversarial Training}
Conditional GAN~\cite{mirza2014conditional} is designed to learn a generative model of a conditional distribution, and shows excellent performance in many tasks, such as image super-resolution~\cite{zhang2020single}. Considering the high correlation between the image and its corresponding text, TGIC uses text features as conditional information in the discriminator for better adversarial training. In the discriminator, the text features are first reshaped, then convolved and upsampled, so that these features can be concatenated with the image. Under the guidance of the text features, the discriminator maps an input ($\hat{x}$,$t$) to the probability $P_{x|t}$.

\begin{figure}[t]
	\begin{center}
		\includegraphics[width=0.95\linewidth]{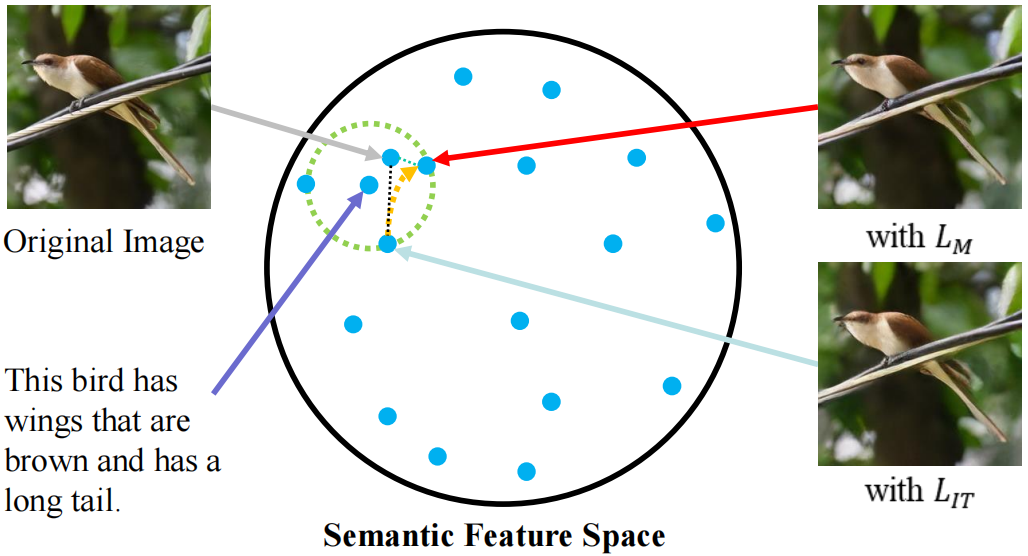}
	\end{center}
	\caption{Projecting multi-modal embedding into the semantic feature space.}
	\label{loss}
\end{figure}

\subsection{Multimodal Semantic-Consistent Loss}
The proposed multimodal semantic-consistent loss is designed to constrain the semantic consistency of the reconstructed image and the original image as well as the text. AttnGAN~\cite{xu2018attngan} suggests to map the image features and text features into a common semantic space with the image encoder and  text encoder, and calculates the negative log posterior probability to make the reconstructed image and the corresponding text semantically consistent. The corresponding loss is defined as
\begin{equation}
	L_{IT}=-(logP(t|f_I(\hat{x}))+logP(f_I(\hat{x})|t)),
\end{equation}
where $f_I(\cdot)$ represents the image encoder.

Although $L_{IT}$ can constrain the semantics of the text description and reconstructions to be consistent, the images that conform to the text description are diverse. Therefore, we also add the constraint between the reconstructions and the uncompressed images, which is shown in Fig. \ref{loss}. The multimodal semantic-consistent loss can be calculated as
\begin{equation}
	L_{M}=L_{IT}+L_{II},~where~L_{II}=\beta||f_I(\hat{x})-f_I(x)||_2,
\end{equation}
where $\beta$ is a hyper-parameter, and $L_{II}$ is the loss function which makes the reconstructions and the uncompressed images semantic-consistent. Specifically, as shown in Fig. \ref{loss}, the text does not contain the description of the bird's eyes, and the reconstruction quality of the eye part in the result generated only by $L_{IT}$ is poor. However, using the improved $L_M$, the subjective quality of the reconstruction result is significantly improved.

In addition to $L_M$, we also use other four loss functions to optimize TGIC. Among them, the reconstruction loss $L_R$, GAN loss $L_G$ and rate loss $L_{Rate}$ are commonly used in the GAN-base codec. Since the text description takes up very few bits and its value is a constant, we do not consider it in the rate loss, but we take it into account when calculating the final bitrates. Here, these three loss functions are defined as
\begin{equation}
	L_{R}= ||\hat{x}-x||_2 ,
\end{equation}
\begin{equation}
	L_G=\mathbb{E}[logf_D(x,t)]+\mathbb{E}[log(1-f_D(\hat{x},t))],
\end{equation}
\begin{equation}
	L_{Rate}=\mathbb{E}[-log_{2}p_{y|z,t}(y|z,t)]+\mathbb{E}[-log_{2}p_{z|\theta}(z|\theta)],
\end{equation}
where  $f_D(\cdot)$ represents the discriminator. Following~\cite{mentzer2020high}, the perceptual loss $L_P$ based on a pretrained AlexNet~\cite{krizhevsky2012imagenet} is also adopted, which is defined as
\begin{equation}
	L_p=||\phi(\hat{x})-\phi(x)||_2,
\end{equation}
where $\phi(\cdot)$ is the function of the pretrained AlexNet.

\begin{figure*}[t]
	\begin{center}
		\includegraphics[width=0.92\linewidth]{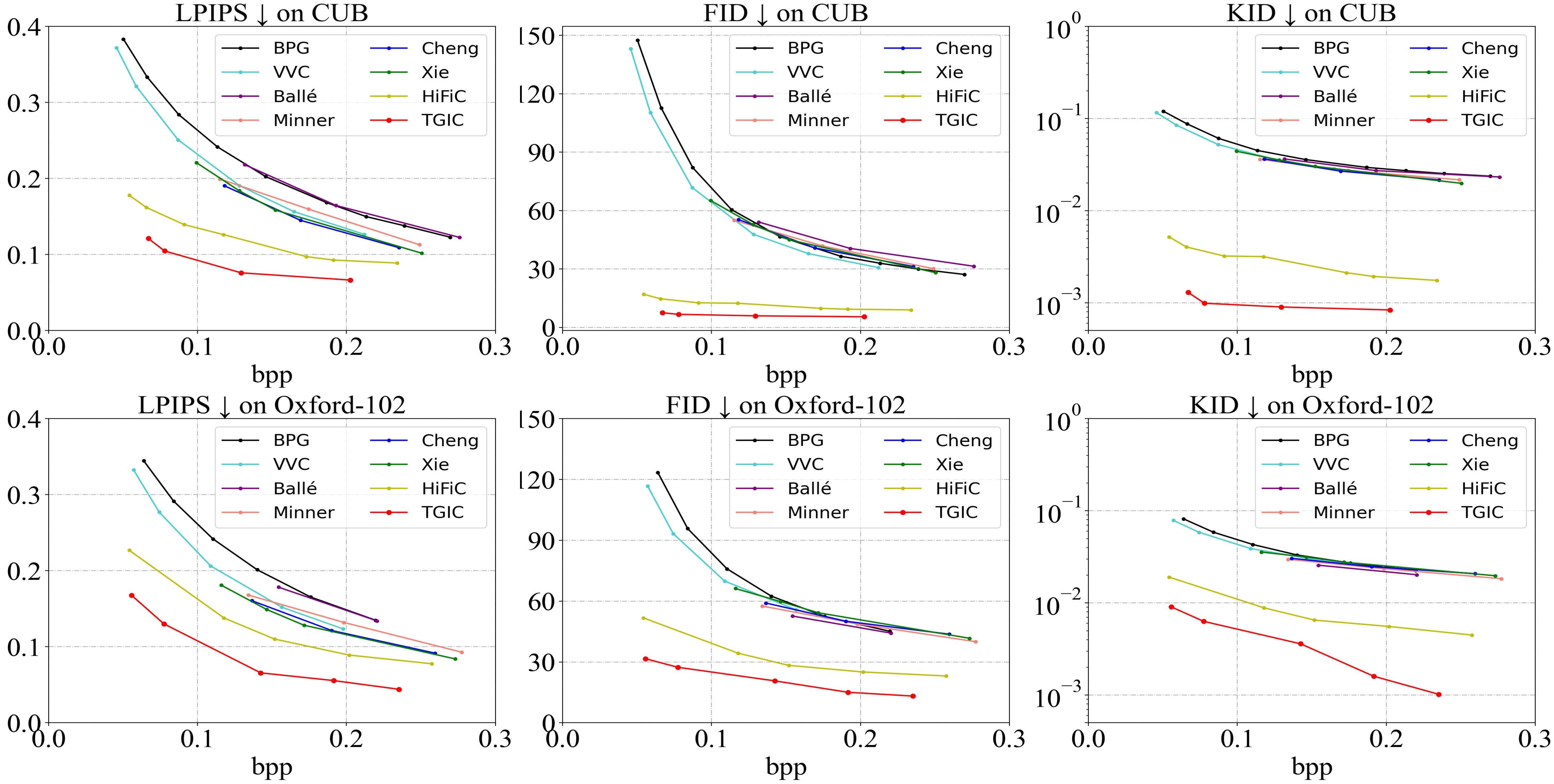}
	\end{center}
	\caption{Performance evaluation on CUB and Oxford-102 datasets in terms of LPIPS, FID and KID.}
	\label{RD-lpips}
\end{figure*}

\begin{figure}[t]
	\begin{center}
		\includegraphics[width=1\linewidth]{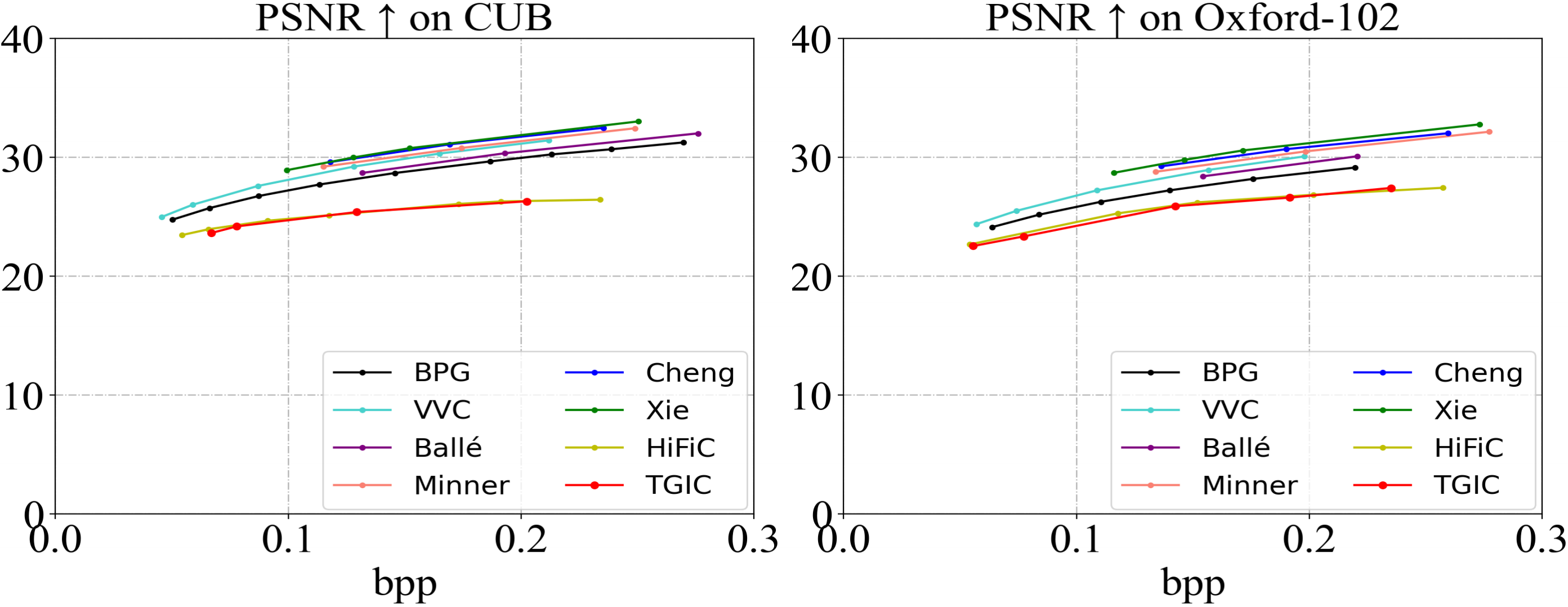}
	\end{center}
	\caption{The PSNR results on CUB and Oxford-102.}
	\label{RD-psnr}
\end{figure}


\begin{figure}[t]
	\begin{center}
		\includegraphics[width=1\linewidth]{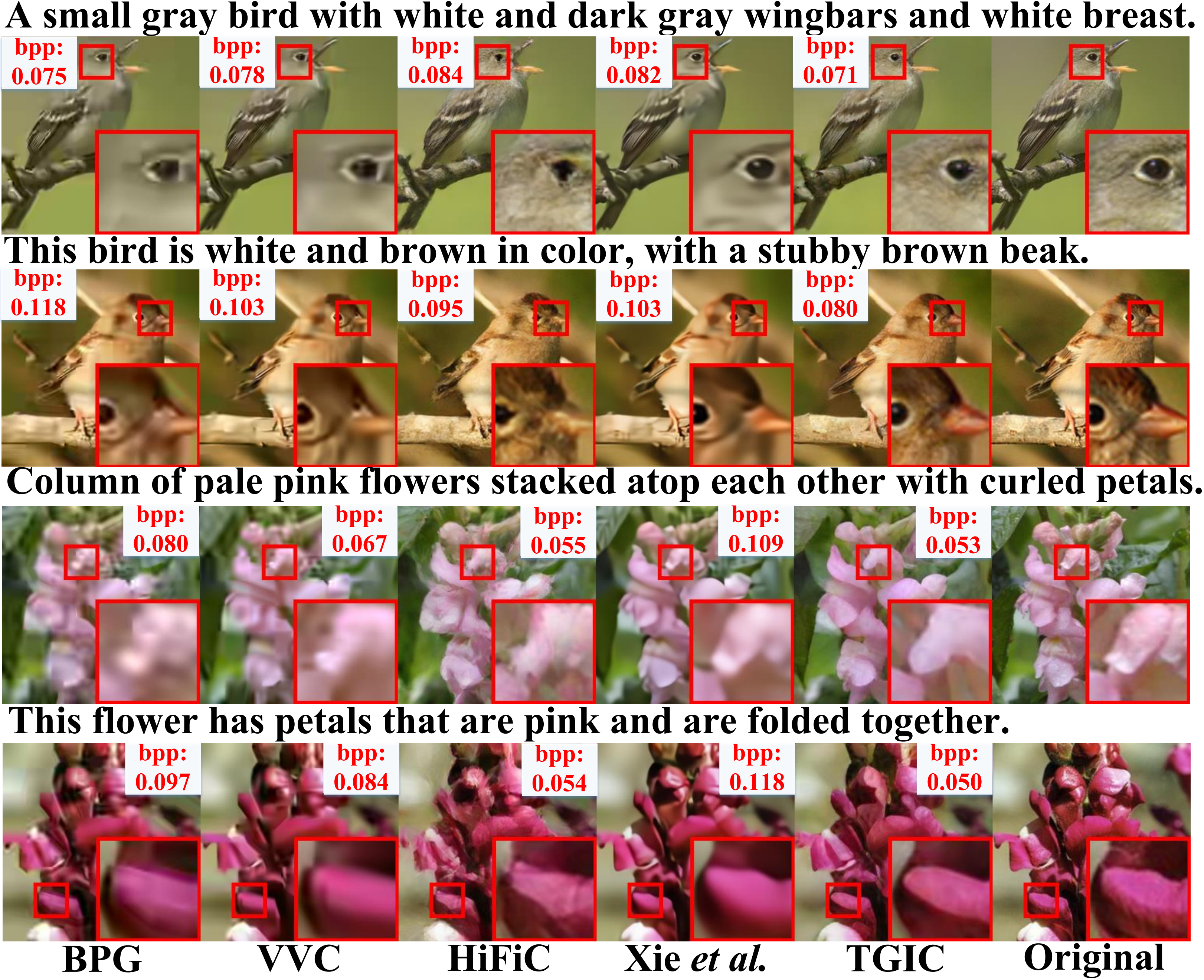}
	\end{center}
	\caption{Visual comparisons with SOTA approaches on CUB and Oxford-102 datasets. Above each line of the images is the corresponding text description. Better zoom in.}
	\label{visual}
\end{figure}

Finally, the global loss function of TGIC is defined as
\begin{equation}
	L=\lambda L_{Rate}+k_{1}L_{R}+k_{2}L_{G}+k_{3}L_{P}+k_{4}L_{M}
\end{equation}
where the $k_{1}$, $k_{2}$, $k_{3}$, and $k_{4}$ are hyper-parameters, and the $\lambda$ is a trade-off parameter to balance rate and distortion. Following~\cite{mentzer2020high}, we adopt the constrained rate. The constrained rate introduces a 'rate target' hyper-parameter $r_t$ and two hyper-parameter including $\lambda_a$ and $\lambda_b$. When the calculated rate is larger than $r_t$, $\lambda$=$\lambda_a$. Otherwise, $\lambda$=$\lambda_b$. We can obtain a model with an average bitrate close to $r_t$ by setting $\lambda_a\gg\lambda_b$

\subsection{Training Implementation}
Following AttnGAN~\cite{xu2018attngan}, the image encoder and text encoder are pretrained to map the image features and text features into a common semantic space. Then the weights of the image encoder and text encoder are fixed while training the proposed TGIC. We adopt Pytorch as the training toolbox, and use the Adam optimization algorithm~\cite{kingma2014adam} with a mini-batch of 4 to optimize the model parameters. All the experiments are conducted on a NVIDIA GeForce RTX 1080 Ti. Our model is optimized for 300 epochs with a learning rate of $1\times10^{-4}$. The hyper-parameters $k_{1}$, $k_{2}$, $k_{3}$, $k_{4}$ and $\beta$ of the global loss function are empirically set as $0.075\times2^{-5}$, 0.15, 5, 0.005 and 40, respectively. The $\lambda_b$ is set as $2^{-4}$, and the $\lambda_a$ is set as ($2^3,~2^2,~2^1$) to adapt to different bitrates.

\section{Experiments}    
\subsection{Datasets and Evaluation}
Following the previous multimodal machine learning-based works~\cite{zhang2017stackgan}, the widely used datasets including CUB~\cite{wah2011caltech} and Oxford-102~\cite{nilsback2008automated} are employed for evaluation. CUB consists of 200 species of bird, with a total of 11,788 images including 8,855 images for training and 2,933 images for testing. Oxford-102 has 102 flower categories, of which 7,034 images are utilized for training and 1,155 images are utilized for testing. The images are resized and cropped into patches of size $256\times256$.

The proposed TGIC aims at improving the codec performance at extremely low bitrates. Thus, we set the rate target below 0.25bpp for TGIC. Note that we use the bit sum of the image and the text to calculate the bitrates for TGIC, while we only utilize the bits of the image to calculate the bitrates for other methods. The bitarate of text are defined as $R_{text}=\frac{Size_{text}\times8}{H\times W}$, where $Size_{text}$ denotes the file size of the text in bytes, $H$ and $W$ denote the hight and width of the image, respectively. Following~\cite{mentzer2020high,yang2021perceptual}, we use LPIPS~\cite{zhang2018unreasonable}, KID~\cite{binkowski2018demystifying}, FID~\cite{heusel2017gans} to evaluate our TGIC and other compared methods, which are highly consistent with human perception of images. In addition, we also use Peak Signal-to-Noise Ratio (PSNR) to measure the fidelity of our results.

\subsection{Comparison against SOTA Methods}
In this part, TGIC is compared with state-of-the-art (SOTA) methods, including BPG~\cite{BPG}, VVC~\cite{VVC}, HiFiC~\cite{mentzer2020high}, and the works of Ball{\'e} \emph{et al.}~\cite{balle2018variational}, Minnen \emph{et al.}~\cite{minnen2018joint}, Cheng \emph{et al.}~\cite{cheng2020learned} and Xie \emph{et al.}~\cite{xie2021enhanced}. Among these methods, BPG and VVC are the traditional methods, and others belong to the deep learning-based methods. Especially, HiFiC is a GAN-based model, aiming to produce results with better subjective quality. We use the BPG software (YCbCr 4:4:4) to test the images. For VVC, we use the VVC Official Test Model VTM 10.0 (YCbCr 4:4:4) of official version with an intra-profile configuration to test on images. We draw the rate-distortion (RD) curves to compare the coding performance of different methods.

\textbf{Quantitative results.} Fig.~\ref{RD-lpips} shows the RD curve comparison on the CUB and Oxford-102 datasets. It can be observed that Our TGIC shows much better performance compared to other methods in terms of LPIPS, FID and KID. Especially, TGIC achieves a comparable or even better performance than other methods, even though these methods are at 2x to 4x bitrates of ours. In addition, we evaluate the fidelity of TGIC in terms of PSNR. As shown in Fig.~\ref{RD-psnr}, TGIC reaches competitive PSNR values compared with HiFiC. This verifies that TGIC can maintain an acceptable fidelity when compressing images towards subjective quality.

\textbf{Qualitative results.}
Fig.~\ref{visual} shows the visualization results of TGIC and other SOTA methods at similar bitrates. It can be found that the results of TGIC show better subjective quality than other algorithms. For VVC and BPG, their results have obvious blocking artifacts. The results of Xie \emph{et al.} are blurry due to the MSE-based training. In addition, HiFiC utilizes GAN to obtain the hallucinated details, and produces visually pleasing reconstructions. Unfortunately, the hallucinated content may be inconsistent with the original image content, resulting in obvious artifacts. Under the guidance of the text description, TGIC can produce satisfied reconstructions with photorealistic details, which are semantic-consistent with the original images.

\begin{figure}[t]
	\begin{center}
		\includegraphics[width=1\linewidth]{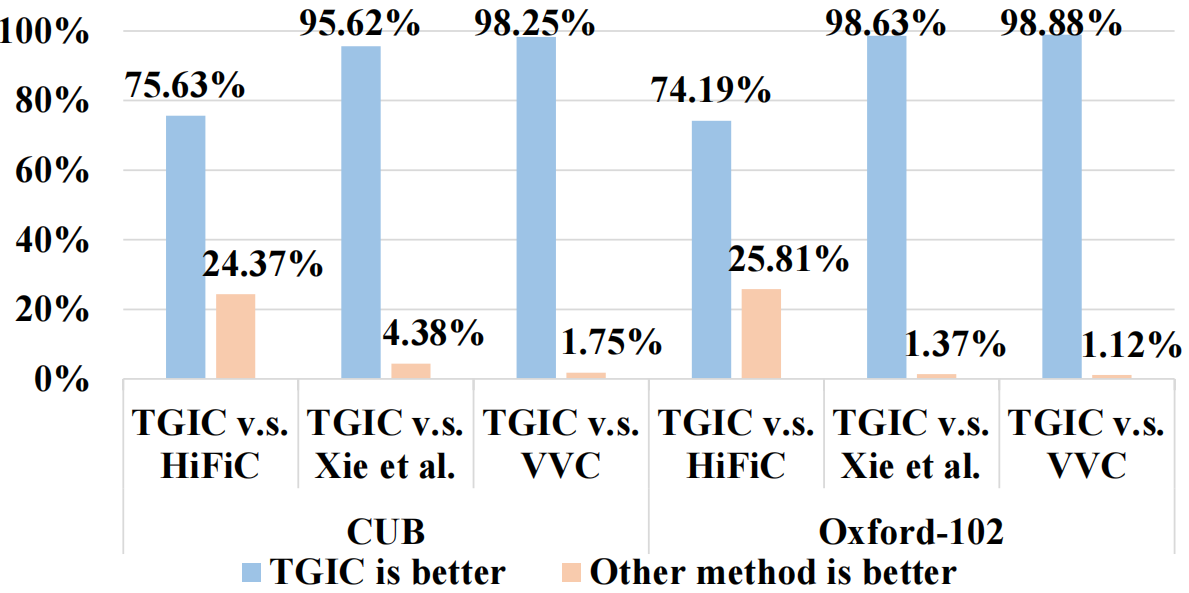}
	\end{center}
	\caption{User study results. The reported value indicates the performance rate of the proposed TGIC against the other SOTA methods, respectively.}
	\label{user}
\end{figure}

\begin{table*}[t]
	\begin{center}
	\resizebox{0.88\textwidth}{!}
	{ 
		\begin{tabular}{c|c|c|c|c|c|c|c|c|c}
			\hline
			\multirow{3}*{Method} &\multirow{2}*{Baseline}&\multicolumn{3}{c|}{Case 1}&\multicolumn{2}{c|}{Case 2}&\multicolumn{2}{c|}{Case 3}&\multirow{2}*{TGIC}\\
			\cline{3-9}
			~ &~&\multirow{1}*{w/o TGFR}&\multirow{1}*{w/o TGIR}&\multirow{1}*{w/o TGAT}&\multirow{1}*{w/o $L_p$}&\multirow{1}*{w/o $L_M$}&\multirow{1}*{w/o IRC}&\multirow{1}*{w/o $L_{II}$}&~\\
			
			~&0.088 bpp&0.077 bpp&0.075 bpp&0.076 bpp&0.077 bpp&0.075 bpp&0.076bpp&0.078 bpp&0.078 bpp\\
			\cline{1-10}
			\hline
			\multirow{1}*{LPIPS}$\downarrow$&0.2411&0.1242&0.1289&0.1174&0.1625&0.1102&0.1297&0.1071&0.1048\\
			\cline{1-10}
			FID$\downarrow$&37.39&8.56&10.45&7.61&7.93&9.09&9.17&9.20&\textbf{6.75}\\
			\cline{1-10}
			KID$\downarrow$ ($10^{-3}$)&18.78&1.68&2.93&1.24&1.46&2.06&2.23&1.77&\textbf{0.99}\\
			\cline{1-10}			
			
		\end{tabular}
	}
	\end{center}
	\caption{Performance Comparisons between variations of our TGIC on CUB dataset. The best results are boldfaced.
	}\label{tab:ablation}
\end{table*}

\textbf{User Study.} 
A user study is further conducted with 32 participants. Given a pair of reconstructed images, the user is asked to judge which one owns a higher perceptual quality and is more consistent with the corresponding text description. We randomly select 50 images from CUB and 50 images from Oxford-102. Note that the bitrates of their compressed results by different methods are required to be similar. TGIC is compared with the state-of-the-art methods including VVC, HiFiC and the work of Xie \emph{et al.}. This user study requires 9600 comparisons in total. Each participant needs to spend 40 minutes to complete the subjective test. As shown in Fig.~\ref{user}, it can be found that the results of TGIC gain more preference than that of other methods, and all exceed 70\%. This subjective comparisons are consistent with the quantitative results in Fig.~\ref{RD-lpips}, which verifies that our TGIC is superior to other algorithms.

\subsection{Ablation Study}

In this part, we study and analyze the contributions of text guidance and different loss functions to our algorithm. Since the resulting bitrates do not exactly match the setting value, we try our best to compare performance of different models at the similar bitrates. We first remove $L_P$, $L_M$ and all the text guidance including text-guided feature representation (TGFR), text-guided image reconstruction (TGIR) and text-guided adversarial training (TGAT) from our TGIC, and regard this model as the baseline model. As shown in Table~\ref{tab:ablation}, our TGIC obtains much better performance than baseline, which confirms the performance gain of introducing text.

\textbf{Case 1: Effectiveness of text.} We conduct the ablation study on the text effectiveness. We test the performances of TGIC without TGFR (w/o TGFR), TGIC without TGIR (w/o TGIR) and TGIC without TGAT (w/o TGAT). As shown in Table~\ref{tab:ablation}, the performances of three models all decrease compared with TGIC. Note that w/o TGIR has the largest performance drop and w/o TGAT has the smallest drop. This means that the text has the greatest effect in the decoder and the least effect in the discriminator.

\begin{figure}[t]
	\begin{center}
		\includegraphics[width=1\linewidth]{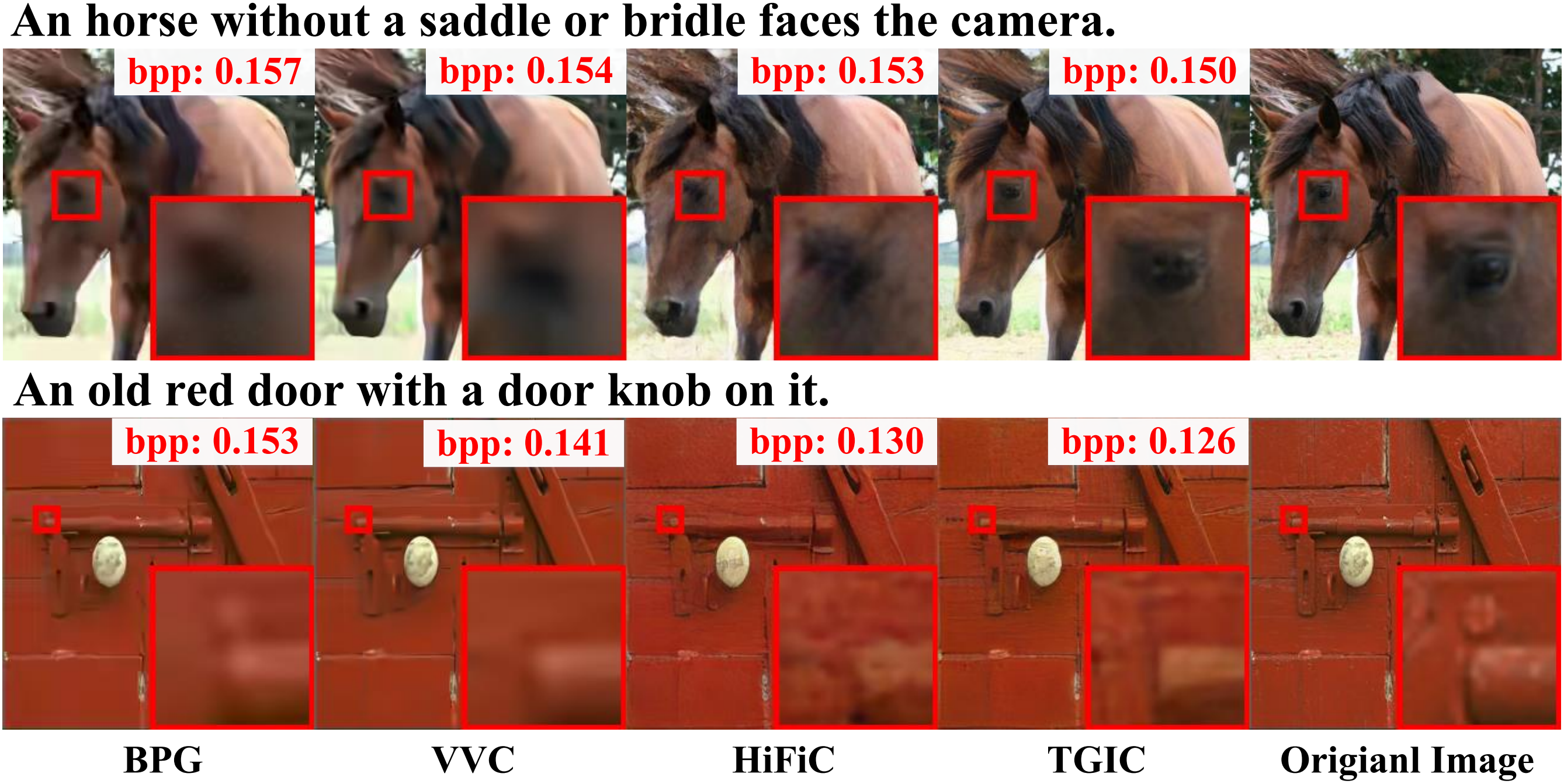}
	\end{center}
	\caption{Visual comparisons with VVC on COCO (first row) and Kodak (second row). Better zoom in.}
	\label{visual_coco}
\end{figure}

\textbf{Case 2: Effectiveness of $L_P$ and $L_M$.} Since $L_R$, $L_{Rate}$ and $L_G$ are common in GAN-based codec, we explore the effectiveness of $L_P$ and $L_M$. As shown in Table~\ref{tab:ablation}, both have performance gains for our TGIC.

\textbf{Case 3: Effectiveness of IRC and $L_{II}$.}
We also take the ablation study to verify the effectiveness of the proposed IRC and $L_{II}$. As shown in Table~\ref{tab:ablation}, the performances both decrease when removing IRC and $L_{II}$. This tells that IRC and $L_{II}$ are benefical to our model.

\begin{table}[t]
\centering
\resizebox{0.47\textwidth}{!}
{ 
		\begin{tabular}{c|c|c|c|c}
			\hline
			\multirow{2}*{Method}&\multicolumn{2}{c|}{COCO}&\multicolumn{2}{c}{Kodak}\\ \cline{2-5}
			~&BPP&LPIPS/FID/KID$\downarrow$&BPP&LPIPS/FID/KID$\downarrow$\\
			\hline
			BPG&0.153&0.289/181.5/0.0426&0.166&0.321/177.0/0.0578      \\
			VVC&0.152&0.250/175.3/0.0418&0.150&0.299/195.3/0.0621\\
			HiFiC&0.140&0.143/110.3/0.0139&0.134&0.142/105.1/0.0198\\
			\hline
			TGIC&0.137&\textbf{0.092/ 77.6 /0.0037}&0.138&\textbf{0.105/ 74.8 /0.0089}\\
			\hline
		\end{tabular}
    }
	\caption{Performance comparisons of our TGIC and VVC on COCO and Kodak datasets. Models are trained on training set of COCO, and tested on Kodak and 1000 images of testing set of COCO. The best results are boldfaced.
}\label{tab:coco}
\end{table}

\subsection{Text-guided Image Compression in Real-World Scenarios}

In real-world scenarios, the image content is diverse and complex. To explore more possibilities of our TGIC, we also conduct experiments on COCO~\cite{lin2014microsoft} with 80 types of objects. To further verify the generation performance, we also conduct the cross-dataset experiments on Kodak~\cite{franzen1999kodak}. Since Kodak does not have corresponding text descriptions, we consider using the image captions methods (\emph{e.g. OFA~\cite{wang2022unifying}}) to generate texts for our experiments. The generated texts are of high quality and highly semantically with the images. We compare our TGIC with BPG, VVC and HiFiC, and the visualization results are shown in Table~\ref{tab:coco} and Fig.~\ref{visual_coco}. We can find that our TGIC produces better results compared to other methods, which confirms the great potential of our TGIC.

\section{Conclusion}
In this paper, we propose a text-guided adversarial generation network for image compression (TGIC). We adopt the image-text attention module to introduce text information into the codec as prior information. Specifically, the text description can help the codec achieve a compact features representation, and can also be used for the image feature enhancement. In addition, we design an image-request complement module to adaptively learn the much-needed guidance knowledge of text information for feature enhancement. Moreover, a new multimodal semantic-consistent loss is well-designed that constrains the semantic consistency between the reconstructions, the texts and the uncompressed images. Experimental results demonstrate that TGIC outperforms the SOTA methods. Especially, even at bitrates below 0.1 bpp, the
TGIC can produce appealing visual results.

\bibliography{aaai23}

\newpage

\appendix
\section{Appendix}

\subsection{The Details of Some Modules}
\begin{figure}[h]
	\begin{center}
		\includegraphics[width=1\linewidth]{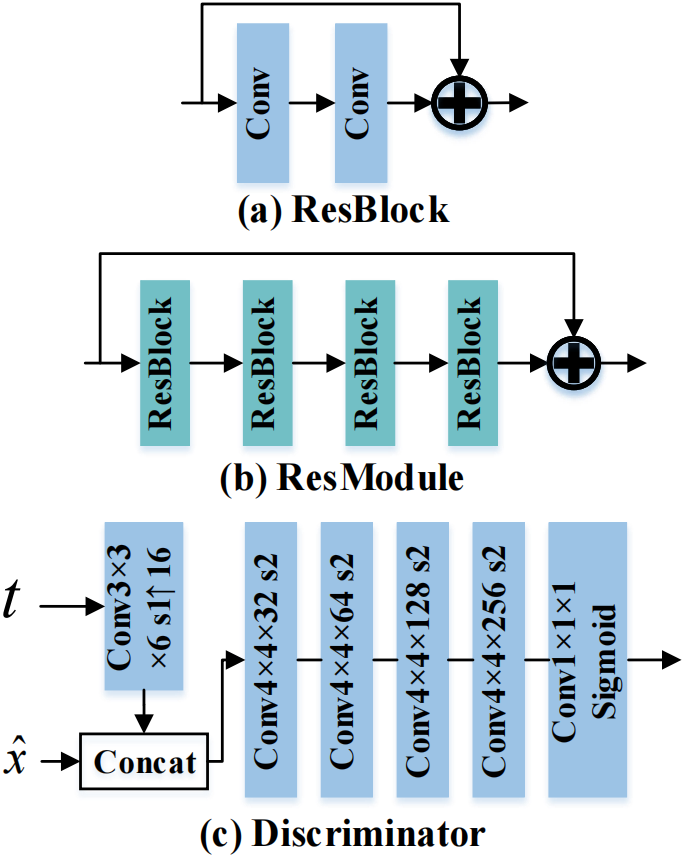}
	\end{center}
	\caption{\textbf{Architectures of the ResBlock, the ResModule and the discriminator.} (a) and (b) show the detail structure of the ResBlock and the ResModule. (c) represents the discriminator. The $Conv3\times3\times6~s1$ means that the convolutional layer of size $3\times3$, stride of 1, and kernel number of 6.}
	\label{fig:module}
\end{figure}

The details of the ResBlock~\cite{he2016deep} in the proposed TGIC are shown in Fig.~\ref{fig:module}(a). The size of the filter is $3\times3$ and the stride is 1. The channel number is the same as the number of the previous convolutional layer. As shown in Fig.~\ref{fig:module}(b), the ResModule consists of four ResBlock modules. The details of the discriminator are shown in Fig.~\ref{fig:module}(c). The $Conv3\times3\times6~s1$ means that the convolutional layer of size $3\times3$, stride of 1, and kernel number of 6. $t$ represents the text features and $\hat{x}$ denotes the image.

\subsection{Visualization Results }

\subsubsection{Results on CUB and Oxford-102 Datasets}

\begin{figure*}[t]
	\begin{center}
		\includegraphics[width=1\linewidth]{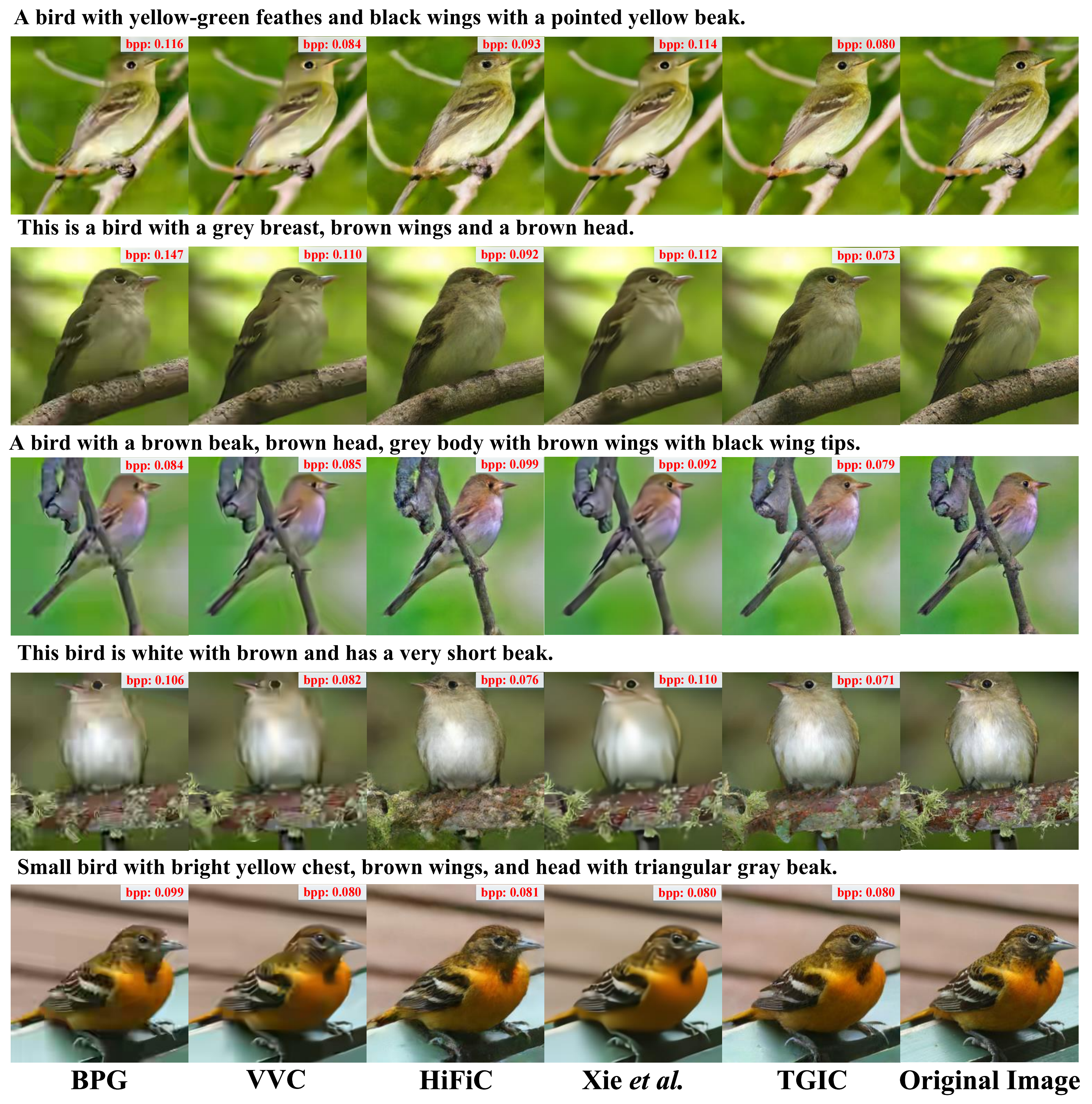}
	\end{center}
	\caption{\textbf{Visual comparisons on CUB dataset of our TGIC and the state-of-the-art methods including BPG, VVC, HiFiC and the work of Xie \emph{et al.}.} The first, second, third, fourth and fifth columns show the reconstructed results of these five methods, respectively. The sixth column shows the original images. Better zoom in.}
	\label{bird}
\end{figure*}

\begin{figure*}[t]
	\begin{center}
		\includegraphics[width=1\linewidth]{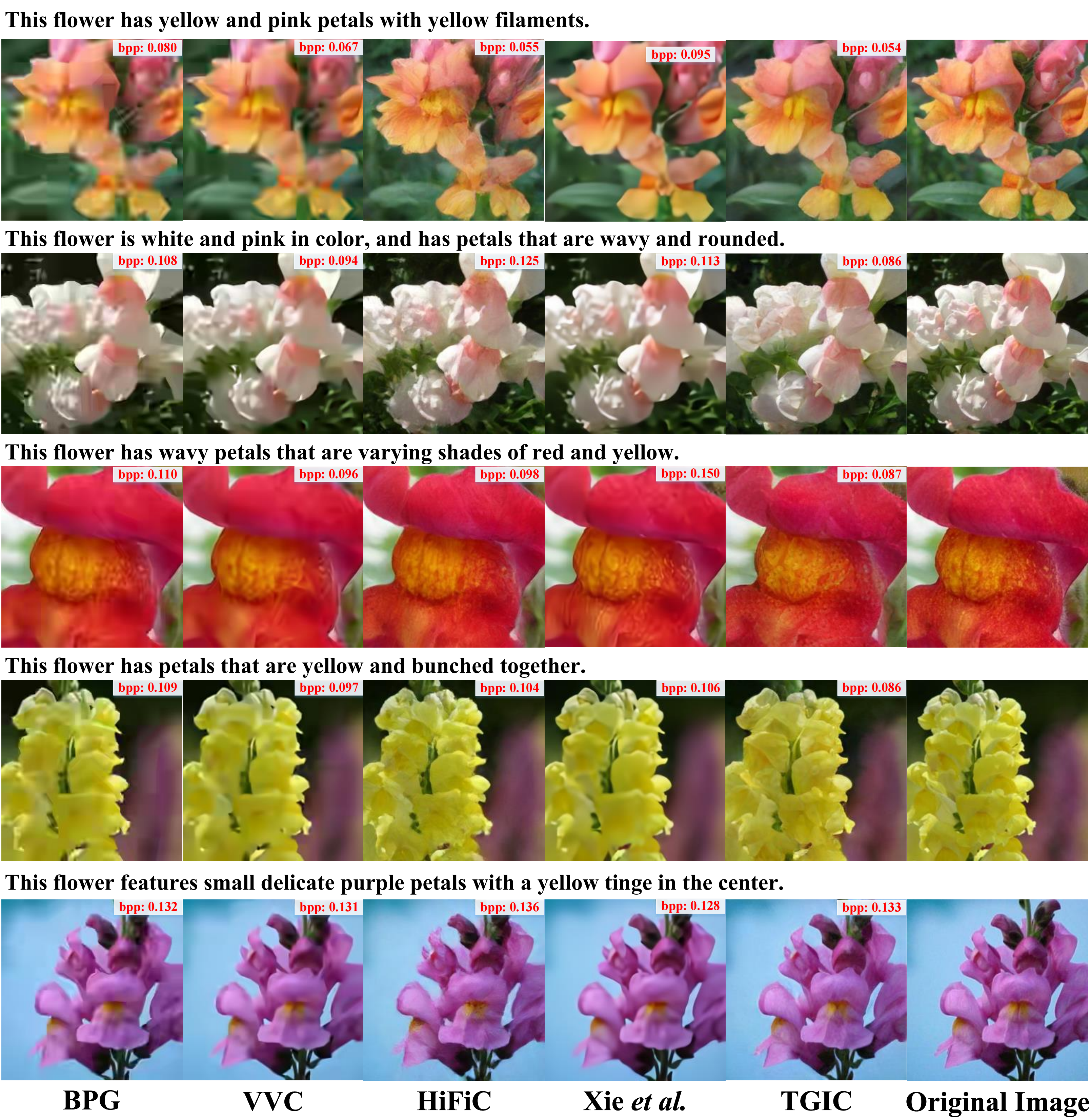}
	\end{center}
	\caption{\textbf{Visual comparisons on Oxford-102 dataset of our TGIC and the state-of-the-art methods including BPG, VVC, HiFiC and the work of Xie \emph{et al.}.} The first, second, third, fourth and fifth columns show the reconstructed results of these five methods, respectively. The sixth column shows the original images. Better zoom in.}
	\label{flower}
\end{figure*}

We add many visualization results on CUB~\cite{wah2011caltech} and Oxford-102~\cite{nilsback2008automated} datasets. The reconstructed results of models, including the proposed TGIC, BPG~\cite{BPG}, VVC~\cite{VVC}, HiFiC~\cite{mentzer2020high} and the work of Xie \emph{et al.}~\cite{xie2021enhanced}, can be seen in Figs.~\ref{bird},~\ref{flower}. In particular, for each image, we adopt these models to encode and decode images at similar bitrates. \textbf{Note that we use the bit sum of the image and the text to calculate the bitrates for TGIC.} As shown in Figs.~\ref{bird},~\ref{flower}, we can find that the results of our TGIC have better subjective quality compared to other methods at similar bitrates.

\subsubsection{Results on COCO Dataset}
\begin{figure*}[t]
	\begin{center}
		\includegraphics[width=1\linewidth]{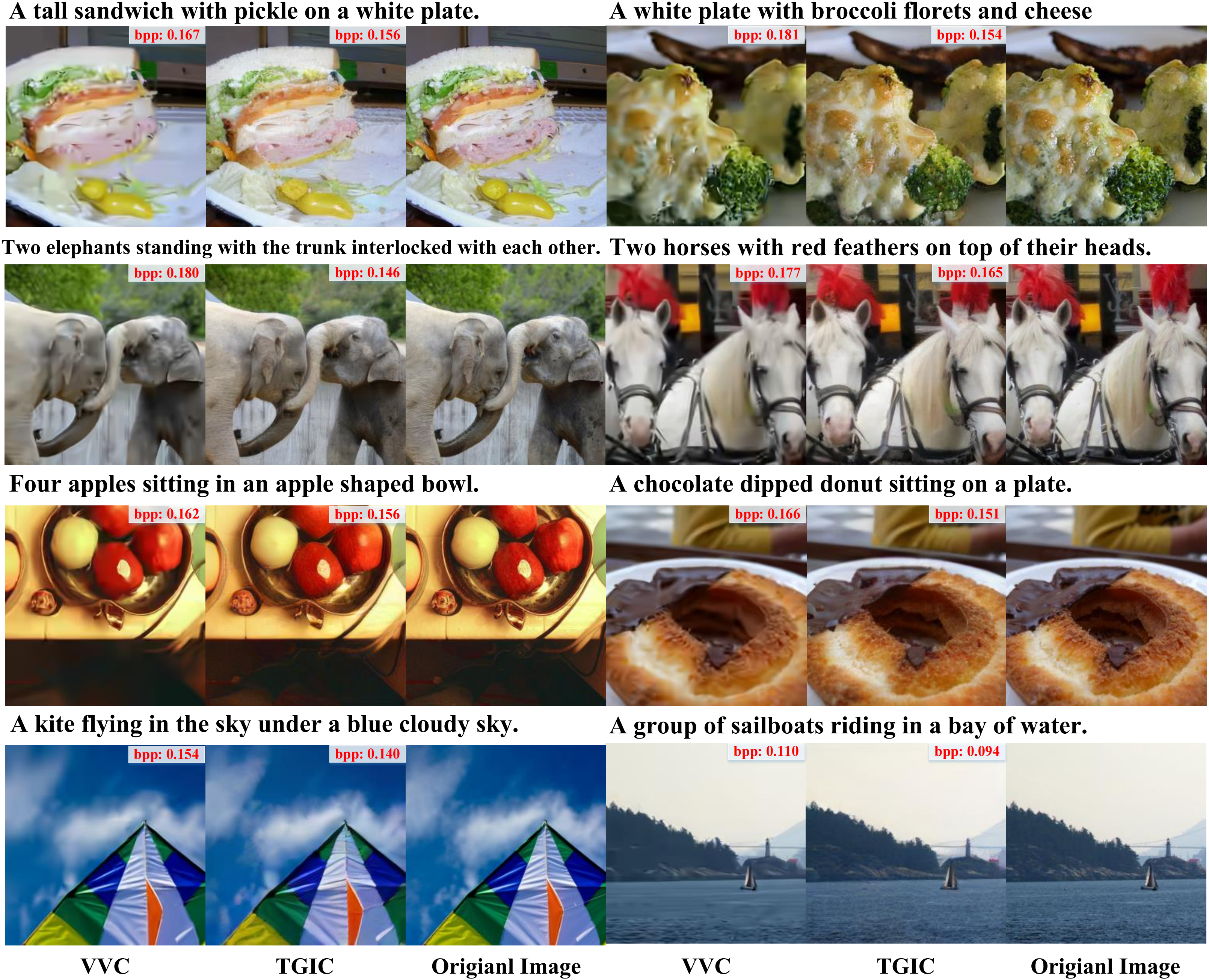}
	\end{center}
	\caption{\textbf{Visual comparisons on COCO dataset of our TGIC and the state-of-the-art method VVC.} The first and fourth columns show the reconstructed results of VVC. The second and fifth columns show the reconstructed results of our TGIC. The third and sixth columns show the original images. Better zoom in.}
	\label{coco}
\end{figure*}

To explore more possibilities of our TGIC, we also conduct experiments on COCO~\cite{lin2014microsoft} datatset, which consists of 80 types of objects. We add a large number of visualization results. The reconstructed results of the proposed TGIC and VVC~\cite{VVC} (the latest commercial codec) can be seen in Fig.~\ref{coco}. We also adopt these models to encode and decode images at similar bitrates for each test image. As shown in Fig.~\ref{coco}, we can find that the subjective qualities of the results generated by our TGIC are higher than that of VVC.

\end{document}